\documentclass[5p,twocolumn,10pt,times]{elsarticle}
\usepackage{amsmath}
\usepackage{caption}
\usepackage{xcolor}

\usepackage{lineno}

\usepackage{multirow}
\usepackage{makecell}
\usepackage{epstopdf}
\usepackage{amsmath}%
\usepackage{amsfonts}%
\usepackage{amssymb}%
\usepackage{graphicx}
\usepackage{mathrsfs}
\usepackage{algorithm}
\usepackage[noend]{algpseudocode}
\usepackage{tikz-cd}
\usepackage{fixltx2e}
\usepackage{url}
\usepackage{arydshln}
\usepackage{fancyhdr}
\usepackage{array}
\usepackage{textcomp}
\usepackage{listings}

\newcommand{\x}[1]{{\color{blue}{#1}}}

\modulolinenumbers[5]

\makeatletter
\def\BState{\State\hskip-\ALG@thistlm}
\makeatother

\usepackage{comment} 
\usepackage{subfig}

\newcolumntype{C}{>{\centering\arraybackslash} m	 } 

\newcommand{\svnidlong}[4]{}%

\usepackage{xcolor}

\definecolor{codegreen}{rgb}{0,0.6,0}
\definecolor{codegray}{rgb}{0.5,0.5,0.5}
\definecolor{codepurple}{rgb}{0.58,0,0.82}
\definecolor{backcolour}{rgb}{0.95,0.95,0.92}
\newcommand{\etal}{et al.}

\newcommand{\naive}{na\"ive}

\begin{document}
\baselineskip11pt

\begin{frontmatter}

\title{Set-based queries for multiscale shape-material modeling}

\author[1]{Oleg Igouchkine}
\ead{oigouchkine@ucdavis.edu}

\author[1,2]{Xingchen Liu\corref{cor1,cor2}}
\ead{xliu@intact-solutions.com}

\cortext[cor1]{Corresponding author}
\cortext[cor2]{Work completed at International Computer Science Institute.}
\address[1]{International Computer Science Institute (ICSI)}
\address[2]{Intact Solutions, Inc.}

\begin{abstract}
 
Multiscale structures are becoming increasingly prevalent in the field of mechanical design. The variety of fine-scale structures and their respective representations results in an interoperability challenge. To address this, a query-based API was recently proposed which allows different representations to be combined across the scales for multiscale structures modeling. The query-based approach is fully parallelizable and has a low memory footprint; however, this architecture requires repeated evaluation of the fine-scale structures locally for each individual query. While this overhead is manageable for simpler fine-scale structures such as parametric lattice structures, it is problematic for structures requiring non-trivial computations, such as Voronoi foam structures.

In this paper, we develop a set-based query that retains the compatibility and usability of the point-based query while leveraging locality between multiple point-based queries to provide a significant speedup and further decrease the memory consumption for common applications, including visualization and slicing for manufacturing planning. We first define the general set-based query that consolidates multiple point-based queries at arbitrary locations. We then implement specialized preprocessing methods for different types of fine-scale structures which are otherwise inefficient with the point-based query. Finally, we apply the set-based query to downstream applications such as ray-casting and slicing, increasing their performance by an order of magnitude. The overall improvements result in the generation and rendering of complex fine-scale structures such as Voronoi foams at interactive frame rates on the CPU.

\end{abstract}

\begin{keyword} Multiscale structure modeling, Multiscale queries, Architected materials, Homogenization
\end{keyword}

\end{frontmatter}


\section{Introduction}
\label{sec:introduction}

\subsection{Motivation}
The application of multiscale structures to mechanical design is becoming increasingly accessible due to the rapid advancement of modern manufacturing techniques. A multitude of methods, algorithms, and tools have emerged to support the design and modeling of such structures~\cite{Antolin2019a, Sitharam2019, Gupta2018, Seepersad2019, Wu2019a}. Periodic lattice and foam-based structures have been applied to architect materials meeting specific material property criteria as needed by the end-user. Commercial tools, such as Autodesk Within and nTopology, have created lightweight infills with these structures. These approaches have specific strengths and weaknesses for addressing specific modeling needs; however, the proliferation of many such methods with specific uses has led to interoperability issues.

A multiscale API based on shape and material queries~\cite{Liu2018a} has been proposed as a possible solution for allowing the integration of different representations across the scales. The architecture allows for rapidly exchanging different fine-scale models, provides a common interface to external tools and downstream applications, and enables easy chaining and linking of multiple scales to create multi-scale structures. In this approach, each coarse- or fine-scale representation is fully encapsulated and only needs to provide methods for a small set of point-based queries for shape or material properties at a given location. 

Downstream applications, such as visualization or slicing for manufacturing planning, are built upon these shape and material queries, which are both fully parallelizable and require a low memory footprint. However, this point-based system architecture requires the repeated and redundant evaluation of local structure within each scale. This computational overhead is manageable when only a little computation is required to generate the fine-scale structure, such as parametrically defined unit cell lattice structures~\cite{Dong2019}. However, it is detrimental to fine-scale structures that require intensive computation, such as sample-based heterogeneous material modeling~\cite{Liu2015a} and Voronoi foam structures~\cite{Martinez2016,Martinez2017}. This repeated evaluation also fundamentally precludes acceleration techniques which could be efficient for groups of points, but are slower on individual queries.

\subsection{Contributions and outline}


We propose a novel set-based query which groups and batch-processes point queries together based on their spatial positions, taking advantage of the spatial locality of the fine-scale structure generation algorithm to accelerate these queries as a whole. This grouping also allows further preprocessing computations which can accelerate the processing of each group but would otherwise be excessive for single-point queries.

The proposed approach differs from simply caching information within a single query, as we not only allow the use of preprocessing which is more efficient on groups of points, but we also sort and re-order computation on the points in order to maximize efficiency. Our approach by itself does not cache or store information between queries, but it is fully compatible with such approaches; we show that in some use cases this can further improve performance.

The concept of set-based queries applies to different fine-scale structures. However, the benefit and performance gain differs depending on the characteristics of the structure. We categorize existing approaches to model fine-scale structures into four quadrants based on the computational resource needed and the geometric similarity of the structure across different neighborhoods. Specifically, we implement two specialized sorting algorithms which are efficient for computationally intensive structures (e.g. Voronoi Foam~\cite{Martinez2016}, inverse homogenization~\cite{Zhu2017}), and for highly-repetitive structures (e.g. parametric lattices~\cite{Dong2019}, and cyclic parametric functions~\cite{Fryazinov2013}). 

The rest of the paper is organized as follows: In Section~\ref{sec:relatedWorks}, we review the existing fine-scale structure modeling algorithm and categorize them based on the metrics we developed. In Section~\ref{sec:methodology}, we formulate the general set-based query and discuss two different point grouping strategies for Voronoi foams and for repetitive structures. In addition, we introduce two different preprossessing algorithms that precompute the Voronoi edges for each neighborhood, and one preprocessing algorithm for repetitive structures. Such preprossessing would not be economic with the point-based query. In Section~\ref{sec:applications}, we examine two downstream applications to our set-based query: ray-casting and slicing. These approaches have fundamentally different types of point sets. While slicing provides a very structured set of points that are fully defined, ray-casting provides a variable set of points due to variable step sizes and early ray termination. In Section~\ref{sec:results}, we test our approach against the \naive{} point-based query on commodity desktop hardware and show that it can improve performance for slicing by an order of magnitude. For ray-casting, our approach can improve performance by over an order of magnitude for the Voronoi foam representation. We also identify the inflection points at which the \naive{} point-based query may be faster for small queries.

Broadly, our work helps to bridge the gap between high-performance, specialized implementations for specific fine-scale structure models and generalized, highly interoperable query-based models. The set-based query lies in the Goldilocks zone between the fully parallel computation of each query point individually, and full precomputation of the entire structure at once. The former approach, as we show, is slower. The latter approach is intractable for multiscale structures, as the level of detail grows exponentially with the number of finer scales, and the entire structure quickly becomes too large to fit in memory.

%
\section{Related Works}

\label{sec:relatedWorks}

\subsection{Fine-scale Structures}
\label{sec:fineScales}

\begin{figure}[htb]
\centering
\includegraphics[width=0.75\linewidth]{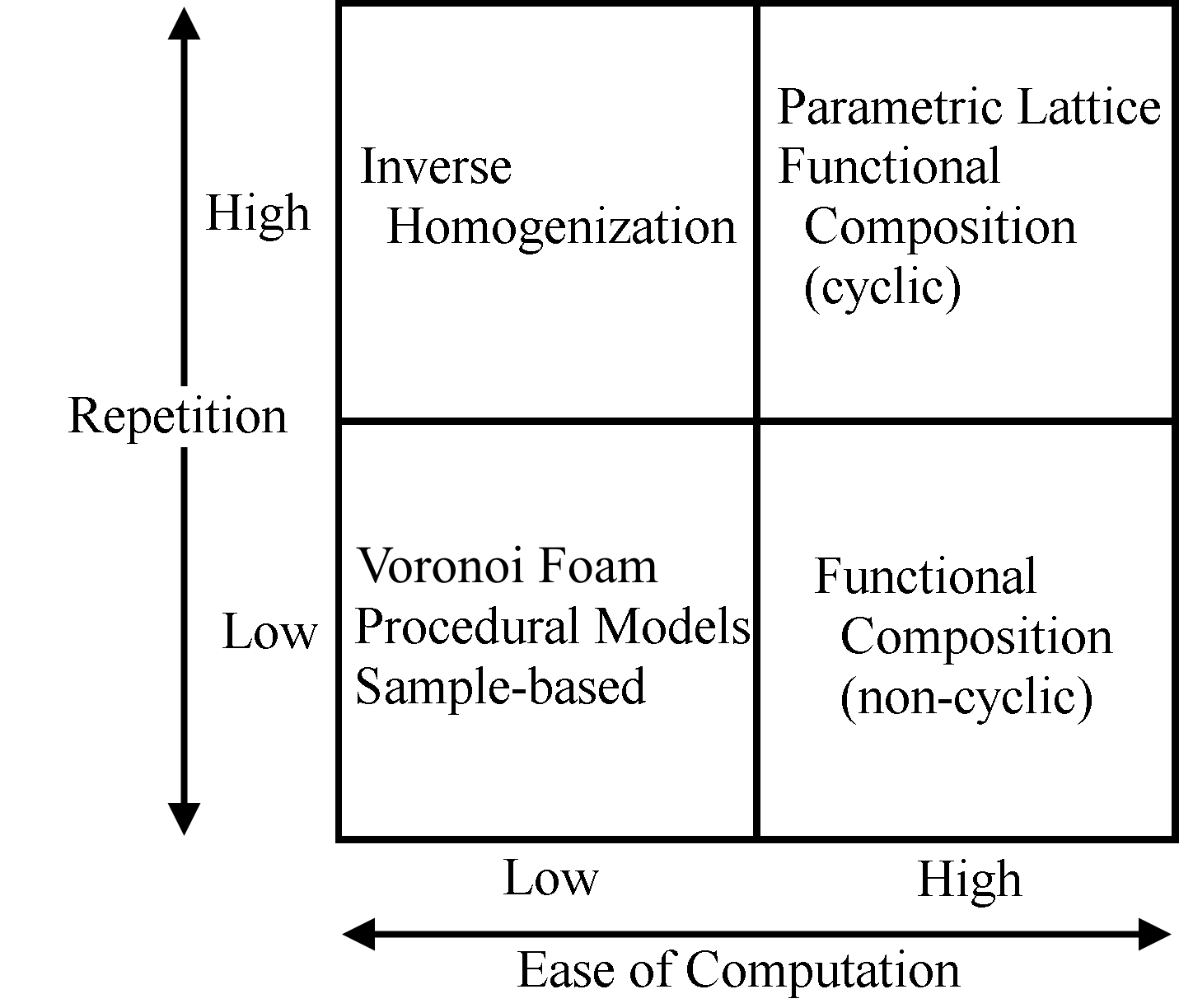}
\caption{A classification of different types of fine scale structures based on the two parameters that affect what can be exploited to achieve a speedup using our approach. Functional representations appear twice, because the repetition of these depends on the particular functions chosen.}
 \label{fig:fineScaleClass}
\end{figure}

The design and representation of fine-scale structures is an active field in which a number of techniques have been recently proposed. Without being exhaustive, we go over a small sample of these. In Figure~\ref{fig:fineScaleClass}, we present a classification of fine-scale structure modeling approaches in terms of computational complexity and geometric repetition. This classification has implications on the grouping and sorting of points for set-based queries, as set-based queries are most helpful where either repetition is high or the cost of computing the structure is high. We choose the parametric lattice~\cite{Dong2019}, a cyclic functional representation, and the Voronoi foam~\cite{Martinez2016} approaches as representatives, as these span the gamut of qualities helpful to efficient set-based queries. We discuss the implications of this further in Section~\ref{sec:methodology}.

\begin{figure}[htb]
\centering
\includegraphics[width=0.95\linewidth]{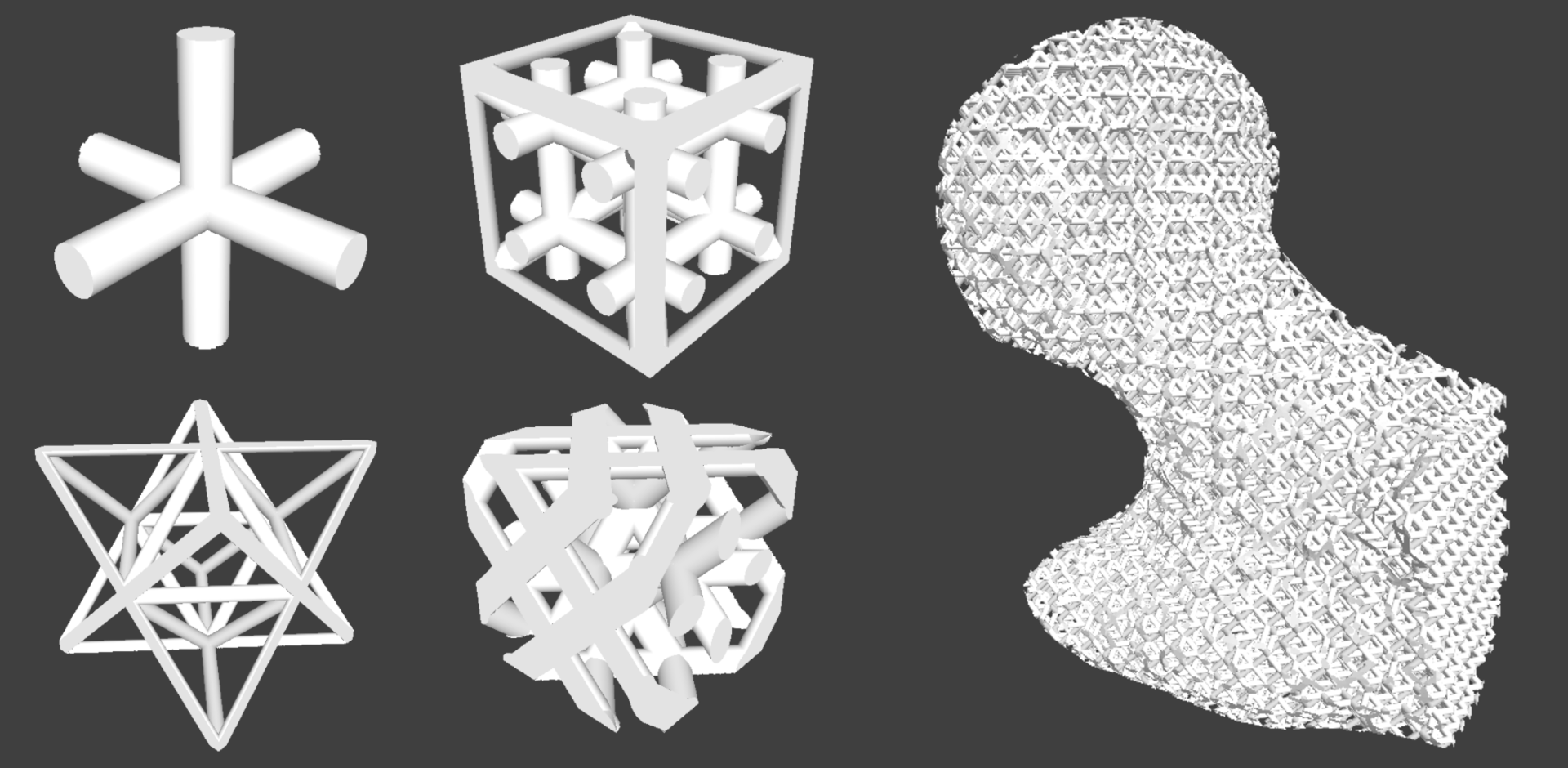}
\caption{Examples of parametric lattice structures. Left: Unit cells which form up a parametric lattice. Right: A coarse-scale femur with a parametric lattice substructure.}
 \label{fig:unitCellEx}
\end{figure}

\paragraph{Parametric Lattices}
The class of parametric lattice structures (Figure~\ref{fig:unitCellEx}) is characterized by the decomposition of the coarse-scale model into a grid of tiled cells, which are filled with structures taken from a precomputed set of unit cells~\cite{Watts2017,Panetta2017}. The exact structure used and any additional parameters are chosen based on optimization to meet the designed material properties required on the coarse-scale, constrained by the possibilities offered by the chosen unit cells and the need for adjacent cells to match up. In Figure~\ref{fig:unitCellEx}, we show some example lattice structures.

\paragraph{Inverse Homogenization} Shape or topology optimization is often used to design unit cell structures with extreme material properties, such as negative Poisson ratio and negative or tunable rate of thermal expansions ~\cite{Sigmund1994b,Larsen1997b,Osanov2016,Sigmund1999,Hopkins2015}. The inverse homogenization process is computationally intensive as multiple finite-element simulations are required to design a unit cell structure. In practice, the inverse homogenization process is often done offline and a surrogate model is constructed to map material properties on the coarse-scale to the detailed fine-scale geometry~\cite{Zhu2017}. One open issue for multiscale structure design with unit cells is the enforcement of the connectivity across the adjacent cells, which is often solved as a global optimization problem~\cite{Zhu2017,du2018connecting}. 

\begin{figure}[htb]
\centering
\includegraphics[width=0.55\linewidth]{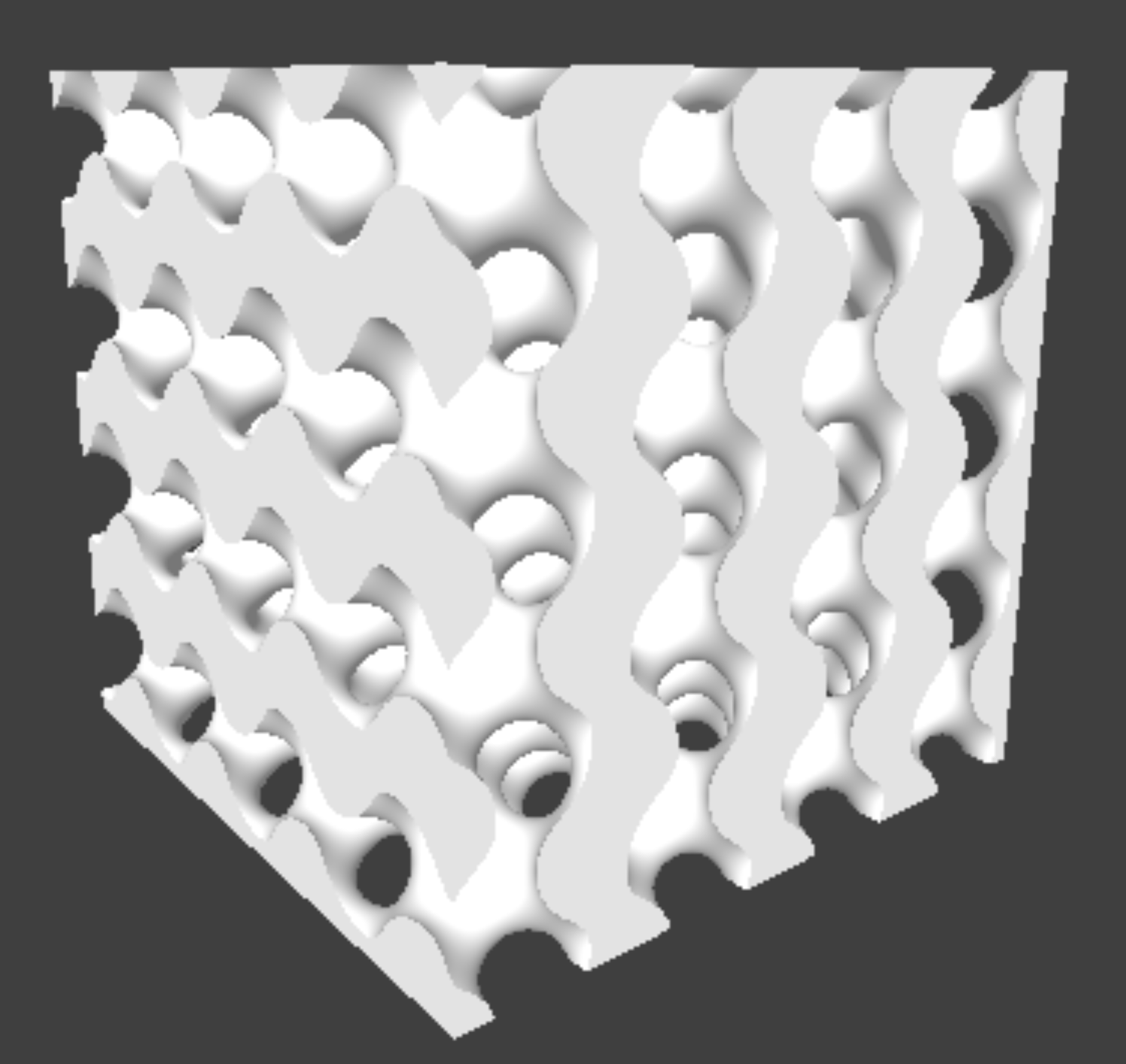}
\caption{An Example of triply periodic minimal surface structures formed with functional representation.}
 \label{fig:funcEx}
\end{figure}

\paragraph{Functional and implicit representation}

Implicit representation model 3D shapes using function with three arguments, e.g. f(x,y,z). A point (x,y,z) is inside the model when f (x,y,z) < 0, outside the model when f (x,y,z) > 0, and on the boundary when f (x,y,z) = 0. The function can be a closed-form expression~\cite{keeter2020massively} or defined discretely as fields~\cite{museth2013vdb}. Functional and implicit representations have been used extensively in the representation of periodic and nearly periodic lattice structures for its compactness and ease of evaluation~\cite{groen2021multi,hong2023implicit}. Figure \ref{fig:funcEx} shows an example of triply periodic minimal surface structures formed with functional representation. Recently, neural networks have been used to represent geometry implicitly, for both complex shapes~\cite{chandrasekhar2021tounn} and lattice structures~\cite{chen2024multi}.

\paragraph{Procedural Models}
The class of microstructures based on procedural models deals with approaches that generate microstructure through procedural functions. Many of these approaches follow a grain-germ architecture, where the shape is generated through the placing of grains (primitive shapes) at germs (locations) and composing them into a microstructure. Examples of such models include Boolean set operations \cite{Stoyan2005}, grain packing subject to non-interference or minimal distance constraints\cite{torquato2002random}, and Voronoi foams.

\begin{figure}
\centering
\includegraphics[width=0.95\linewidth]{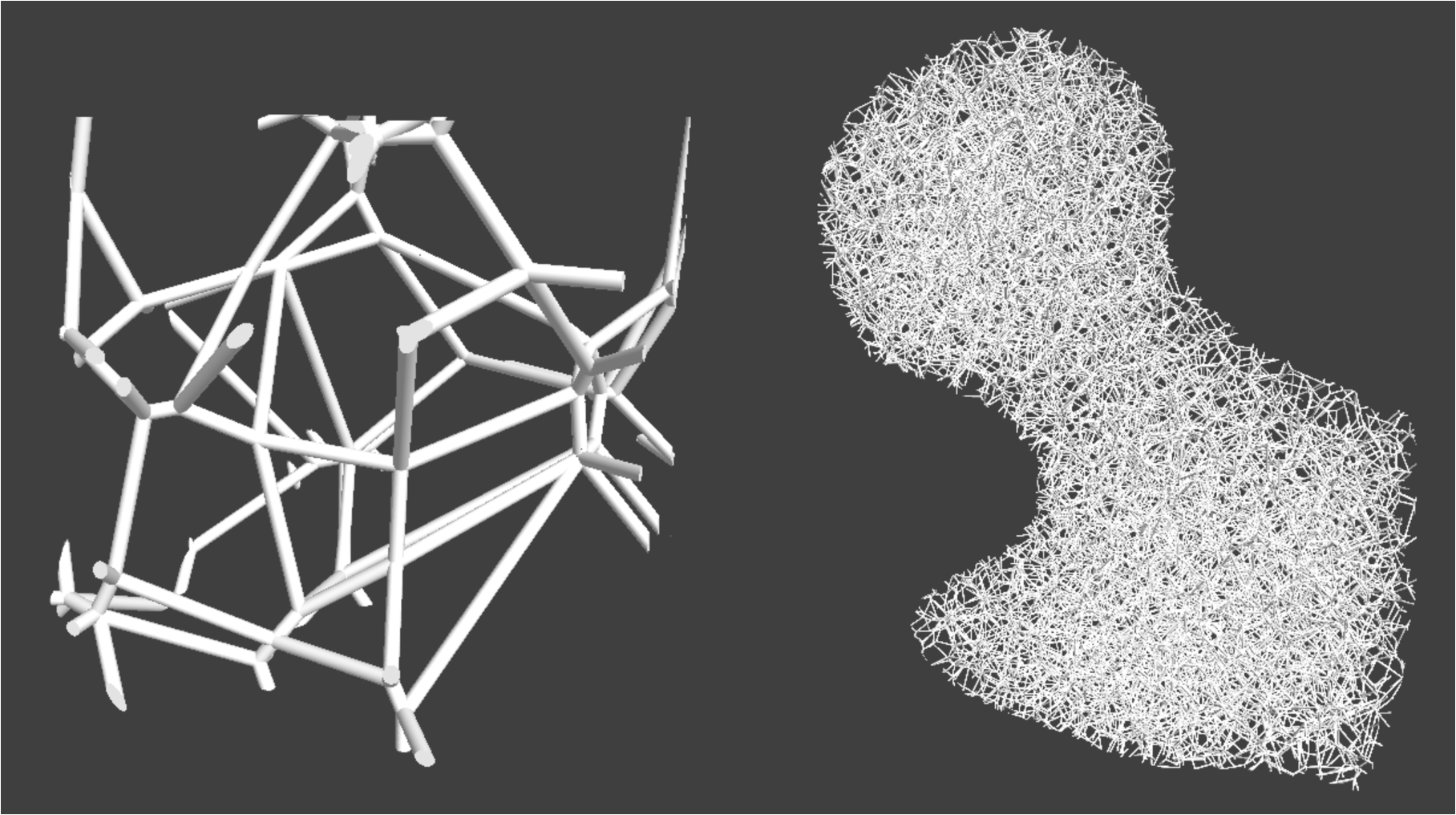}
\caption{Examples of the Voronoi foam structures.}
 \label{fig:voronoiFoamEx}
\end{figure}

\paragraph{Voronoi Foams}
The Voronoi foam (Figure \ref{fig:voronoiFoamEx}) is another popular  procedural, aperiodic model for the purpose of easily realizing gradient material properties \cite{Martinez2016,Martinez2017}. The basic structure is determined by a set of seed points and the Voronoi diagram defined by them. By ensuring a regular distribution of seed points, one seed point per neighborhood cell, the set of seed points required to characterize the foam at a given location remains local and gives a constant-time algorithm for representing the foam. The geometrical properties of the Voronoi diagram ensure that the Voronoi foam is globally connected, and the distribution of the seed points allows for easy manipulation of the physical properties at the macroscale.

The basic algorithm for determining whether a given point $p$ is inside or outside the foam consists of first finding the finite set of seeds $S$ from the neighborhood; geometric properties limit our search size to the local area. The closest seed point $s_c$ determines the cell in which $p$ is located, and every triplet $\{s_c, s_1, s_2\}$ of seed points including $s_c$ is then checked to see whether $p$ is within the cylinder described by $\{s_c, s_1, s_2\}$; a final check is then performed to ensure that this cylinder indeed forms part of the Voronoi cell around $s_c$.

\paragraph{Sample-based Random Heterogeneous Materials}
Sample-based methods allow the reconstruction of random heterogeneous material structures through texture synthesis applied to samples of existing material. The sample-based approach can reconstruct materials with guarantees on common material descriptors, such as correlation functions and Minkowski functionals, and on the effective material properties~\cite{Liu2015a}. Bostanabad et al~\cite{Bostanabad2016a} improve the speed of the reconstruction through feature-based supervised machine learning. The sample-based approach has also been applied to design functionally graded materials and anisotropic material structures~\cite{Liu2017,Liu2017b}. One unique feature of the sample-based approach is that the coarse-scale material properties are mapped to the fine-scale structures with overlapping neighborhoods, allowing a higher sampling rate but increasing the computational cost of this approach. Recently, deep learning with feature learning capability has been applied to increase the computational efficiency of the sample-based approach~\cite{Bostanabad2020,Cang2017,chen2021geometry}.

\subsection{Multiscale Shape-Material Modeling API}

\begin{figure*}[htb]
\centering
\includegraphics[width=0.95\linewidth]{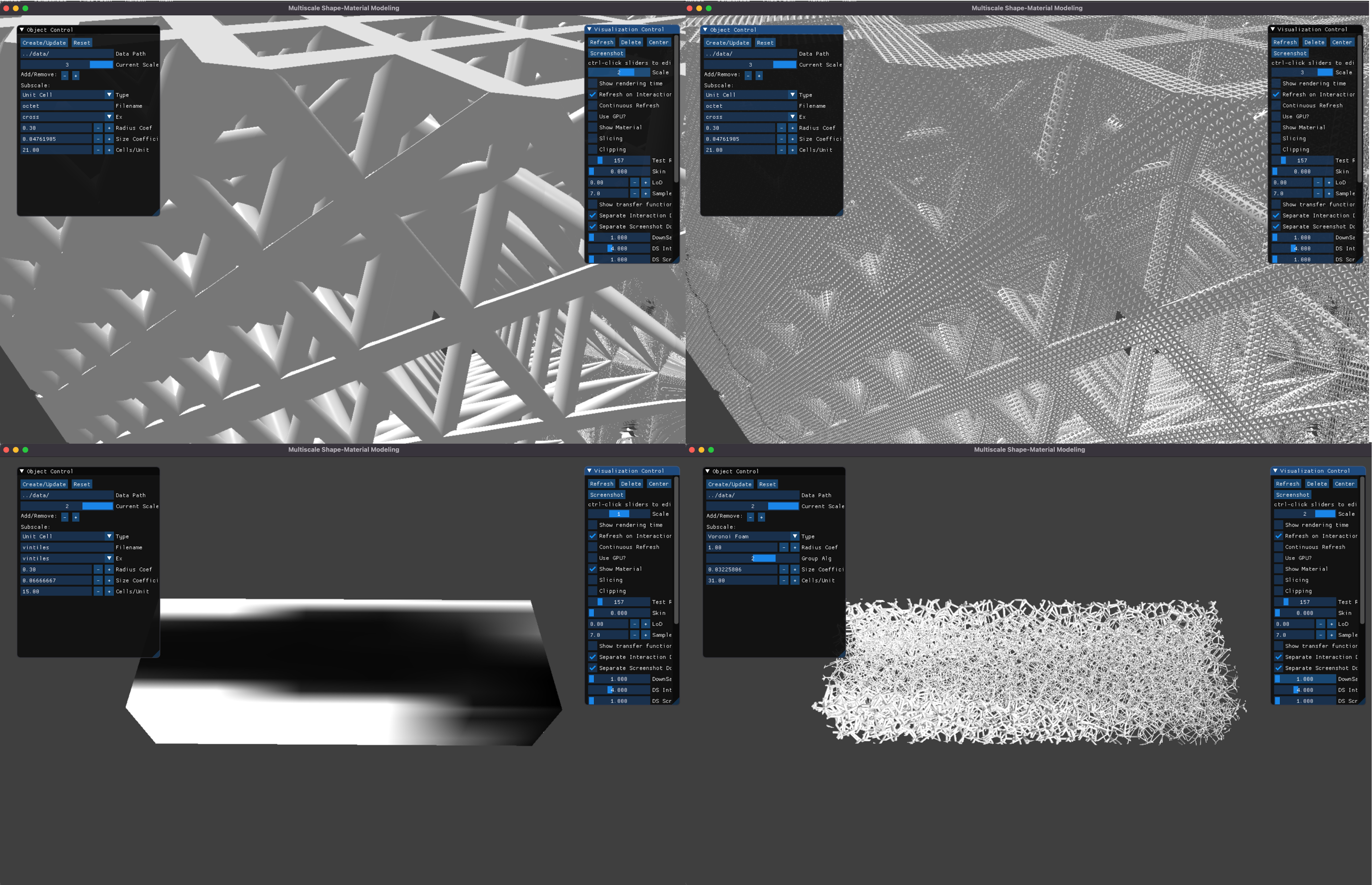}
\caption{An illustration of the multiscale shape-material modeling API. Top: A three-scale structure with parametric lattices visualized at the second and third scales. Bottom: A two-scale cantilever beam with functionally graded Voronoi foam structures on the fine scale following the effective material property field on the coarse scale.
}
\label{fig:multiscaleEx}
\end{figure*}

The wide variety of different fine-scale methods creates an interoperability challenge, especially when attempting to use multi-level structures. 
A query-based approach to multiscale shape-material modeling was proposed by Liu and Shapiro~\cite{Liu2018a, liu2021application} and serves as the platform for our current investigation. The fundamental model is that a series of representations, from coarse-scale model to finer-scale details, are combined to give a final shape and material function for the overall structure. The fine-scale structure modeling approaches reviewed in the first part of this section are abstracted as downscaling functions. Adjacent scales are related to each other through these downscaling functions, which ensure that material properties are consistent and adhere to constraints placed upon the coarse model. Multiscale models are represented through repeated application of downscaling functions. Figure~\ref{fig:multiscaleEx} shows an example of a 3-scale model as modeled by this approach; the finest scale is only visible upon close interaction. The material property field on the coarse scale may be designed through material optimization~\cite{Watts2019,Liu2019b,zhang2020stress}.

Each fine-scale model is defined through a combination of the neighbourhood, which defines grid spacing, and the specific representation of the microscale structure within each neighbourhood cell.
This can be intuitively understood through the example of the unit cell microstructure, where the microscale structure might be a 3D cross of rods and the neighborhood may be a repeating grid of cubes; the combination of these two forms a lattice structure over the coarse model. We discuss additional examples in the following sections.

Queries to the multiscale model are given at a certain length scale, which determines at what scale the query stops and ensures that the result is given at the appropriate level of detail. Specifically, the API requires three fundamental queries from each scale~\cite{Hoffmann2014}: point-membership, which defines whether point p is inside or outside the structure; distance, which determines the distance from p to the closest contour; and material, which determines what the material property is at p for a given length scale. The implementation details on the ray casting algorithm for visualization using these fundamental queries are overviewed in Section \ref{sec:raycasting}. Interested readers may refer~\cite{liu2021application} for an in-depth discussion.

\section{Set-based queries}
\label{sec:methodology}

\begin{figure}[htb]
\centering
\includegraphics[width=\linewidth]{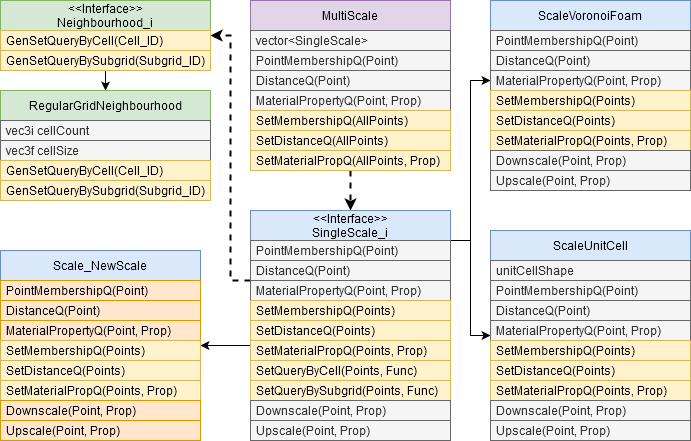}
\caption{The class diagram for our multiscale API. Functions and members in gray exist with the existing multiscale API; functions in yellow represent our proposed additions. Only the orange functions within Scale\_NewScale must be provided when adding a new scale, while those in yellow may also be added for additional performance.}
 \label{fig:classDiag}
\end{figure}

\subsection{Basics of Set-based Queries} 
The generalized set query consists of answering the point-membership, distance, or material queries for a discrete set of multiple points simultaneously. Our methods support both structured and unstructured sets of points as input. However, set-based queries with structured sets can be further optimized, which will be discussed in Section~\ref{sec:applications}. Similarly to point-based queries, each set-based query is answered on each scale of the multiscale structure independently, starting from the coarsest and moving down to the finest. On each scale, points that are not inside the model are either answered trivially (point-membership, material query) or using information from the current scale (distance query), then are removed from further consideration. Interested readers may refer to Liu \etal~\cite{Liu2018a,liu2021application} for the detailed formulation of the point-based multiscale queries.

\begin{figure}[htb]
\centering
\includegraphics[width=\linewidth]{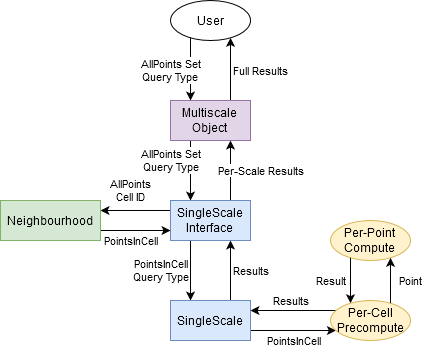}
\caption{A flowchart of how data flows during a set query. The user describes where and what to compute through Set\ldots{}Q, which the multiscale object dispatches to each individual scale to compose a result. Each scale uses SetQueryBy\ldots to break the set of all points into smaller groups; the neighbourhood takes care of generating or grouping points. These groups are then passed to the Set\dots{}Q function of the individual scale, which first does some pre-computation over the group before computing values per-point. If a scale author doesn't provide a specialized function for Set\dots{}Q, it will simply call the individual point query for each point in a group.}
\label{fig:dataDiag}
\end{figure}

In Figure~\ref{fig:classDiag} we show a class diagram of the overall multiscale API enhanced with set queries. A multiscale object holds all scales comprising the overall model and is in charge of dispatching calls to each individual scale. For a set membership query, the user calls Multiscale's SetMembershipQ with a description of all desired points. Each individual scale will take the set of all points and group them according to their needs by calling the provided SetQuery functions, which will generate groups of points through functions in the Neighbourhood definition. Each group of points is then processed in parallel with calls to SetMembershipQ within the individual scales. If SetMembershipQ is not provided by the authors of a given scale, then a default implementation runs point queries overall points in a group-by-group manner. Figure~\ref{fig:dataDiag} shows this flow of data through the classes comprising a multiscale object.

The description of sample points provided by the user may simply be an array of all desired point query locations. However, for certain applications, such as slicing, the desired point locations can be described parametrically. In these cases, the user can pass the parameters to the MultiScale class, and SetQueryBy\ldots will generate the points for each group on the fly, thus avoiding the overhead associated with identifying which group pre-generated points are associated with.

Each set query consists of two components: first grouping (or generating) points by the ability to share pre-processing (SetQuery, GenSetQuery), then processing each group of points in parallel (SetMembershipQ, SetDistanceQ, SetMaterialPropQ). The obvious approach to group points is by local neighborhood cell, and this approach is recommended for the general case; however, other groupings may work better for certain downscaling functions, such as grouping queries by the local subgrid \textit{within} a neighborhood. We implement both types of grouping as a basic function for our downscaling function, allowing authors of new downscaling functions to easily group the points in a set query according to their needs. 
While we generally expect that highly-repetitive structures such as parametric lattices will benefit more from this subgrid grouping, our tests have shown that even for such structures the neighbourhood grouping is faster.
We remind the reader that the neighbourhood cell is the local spatial region (grid cell) of a single fine-scale structure.

The processing of each group of points in parallel generally requires some type of precomputation to achieve a significant speedup, and this generally is specific to the downscaling function being used. This creates a critical amount of points that must be in each group to achieve speedup over the point-by-point approach. We calculate these critical points in Section~\ref{sec:results}, and they can be used as a guide for switching between a point-based and set-based approach.

\subsection{Preprocessing for Voronoi Foams} 
\label{sec:setQVF}

Our first use case for the set query is the Voronoi Foam representation presented in Section~\ref{sec:fineScales}. In this case, we use a neighborhood-based grouping and apply special preprocessing to each neighborhood cell. First, we can generally speed up the computation by discarding seed points which are guaranteed not to contribute to the Voronoi foam at a given query location $p$; we can characterize the set of necessary seed points as the natural neighbors of $p$~\cite{Sibson1980a}, however, the act of finding the natural neighbors for each query point is generally more expensive than the speedup gained by discarding extra seed points. 

However, in the case of the set-based query, we can do this efficiently by first finding the Voronoi diagram, or equivalently the Delaunay tetrahedralization (DT), for an entire neighborhood cell $N$. The same structure may be used across an entire neighborhood because the greater set of seed points is always the same within each neighborhood, while at the same time we retain efficiency by constraining the actual $O(n*lg(n))$ construction of the structure to the finite set of local seed points. We use the Bowyer-Watson algorithm to generate the DT. In order to avoid floating-point precision errors, we re-scale all seed points to the 0-1 unit cube before calculating the DT, then re-scale all computed circumspheres back to the original position. If there is not a sufficient amount of points per neighbourhood, this precomputation will cause the overall performance to decrease; we experimentally derive the values in Section~\ref{sec:sliceres}.

We further speed up the process by precomputing Voronoi foam edges for each grid cell, which shifts the burden of computation into the per-cell precomputation from the per-point computation. We can do this either by adapting the algorithm used by Martínez et al.~\cite{Martinez2016} or by pre-generating the Voronoi foam for the given cell. We implement and test both these approaches in Section~\ref{sec:results}.

\paragraph{Method 1} 
In the first case, we create edges from every group of 3 seed points and test for whether we are in the correct Voronoi cell on a per-sample-point basis. These are created in exactly the same way as in the point-by-point sampling case and simply allow reuse of computation on a per-cell basis. We optimize this somewhat further by discarding edges that do not intersect the current grid cell.
    
\paragraph{Method 2}
In the second case, we instead find all pairs of adjacent tetrahedra in the Delaunay tetrahedralization and create edges between the circumcenters of each pair. We also store the IDs of the three seed points forming the shared face. Since we don't use adjacency information in our construction, we must find all pairs by testing whether each pair of tetrahedra shares a face explicitly; among the different approaches we tried, we found the fastest way was to iterate over all tetrahedra once, store unmatched faces in a list, and compare all new tetrahedra to all faces in the existing list.

When testing against the Voronoi foam, both in the point-based and set-based queries, we can accelerate the testing by finding the closest seed point to the sample position and only testing the Voronoi edges formed by seed points including the closest. Storing the IDs associated with each edge while pre-computing them allows us to use this acceleration with the set-based query. Compared to the original approach presented in~\cite{Martinez2016}, we introduce the per-cell overhead of constructing the Delaunay tetrahedralization, precomputing Voronoi edges, and discarding unnecessary edges. However, the per-point query is greatly accelerated as we no longer need to compute Voronoi edges for each point, and because we reduce the number of Voronoi edges to test against.

\subsection{Preprocessing for Repeating Structures}
\label{sec:paramstruct}

Our second use case is the class of highly-repetitive fine-scale structures. 
We create an approach that can be generally used on any structure in this class. In general, for our approach to work, we assume that each point can be transformed into a local position within a cell by the neighborhood function and that the geometry within the unit cell depends only on a single parameter $r$. We envision that at any sample point in a cell, the signed distance function monotonically decreases with increasing $r$. It is not necessary that the value of $r$ be constant between different cells, and this could be extended to multiple types of unit cells at once by operating on each unique geometry in parallel.

Our approach subdivides the local neighborhood into $n^3$ local clusters set up on a grid. 
We efficiently speed up the distance and membership queries by then finding the interval of values of $r$ for which all corner points of the cluster are inside or outside the structure. 
We then determine the value of $r$ for every point in the local cluster and test whether this is within the interval of that cluster; if we are outside of this range, we can immediately return true or false for the point-membership query without testing the individual point against the geometry of the unit cell, and immediately return an approximate distance for the distance query where appropriate.

The interval of values for $r$ can be analyitically found for parametric lattices by first finding the minimal distance between the center of each cylinder forming the lattice and each cluster's corner points, then finding the minimum and maximum of these; for the general case, we instead need to sample within each subgrid. For parametric lattices, we also perform an intersection test to ensure none of the cylinder centers pass through the subcell, in which case we set the minimum $r$ to $0$.

This acceleration strategy may be used both when sorting points by neighbourhood and by subgrid cell, as we find the first case to be faster. To do this, we precompute the subgrid $r$ values, which only depend on the common unit cell geometry, then do the testing within each neighbourhood.

\section{Application of the set-based queries}
\label{sec:applications}

Downstream applications may require either structured or unstructured sample locations, influencing how these points are presented to the set-based query algorithm. We explore two specific applications which span this gamut: slicing and ray-casting. Slicing our multiscale model uses a highly-structured grid of points that may be precomputed, while ray-casting operates on an unstructured set which changes while the algorithm is running.

\subsection{Structured and Unstructured Set Queries}

The grouping of sample points by spatial position can be done either through general approaches for grouping arbitrary points or through specialized techniques for structured queries. These can be categorized as approaches that are point-first or set-first; in the former, all sample positions are explicitly given and we sort each point into a group, while in the latter we generate the points needed for each group as we go. Because sorting points is an expensive operation, it is preferred to generate the points wherever possible. We implement both for common grouping algorithms in our interface, which can be used by authors of new microscale representations.

\paragraph{Arbitrary Points}
For entirely arbitrary point sets, it is necessary to evaluate which group each point belongs to, and we cannot enumerate all points within a group without first doing this. Therefore, the grouping of points is done by iterating through all points and finding the group to which they belong, ensuring that we only iterate through each point once, and producing a list of points for each group. We then present each group's list to the set query implementation of each microscale. The downside of this approach is that the sorting step is slow, and is either not easily parallelizable, or does not benefit from it on the hardware we tested.

\paragraph{Regular Grids}
We can represent regular grids of points parametrically by the number of points and their spacing, mapping to the output by the position of the point. For by-neighborhood-cell sorting, we can then generate the points on a cell-by-cell basis by finding the first point within a given neighborhood, then stepping from there; this avoids the sorting process, automatically generating points into memory-contiguous arrays, and allows full parallelization of the grouping step. A similar approach is used for by-subgrid sorting.

\begin{figure}[htb]
\centering
\includegraphics[width=0.95\linewidth]{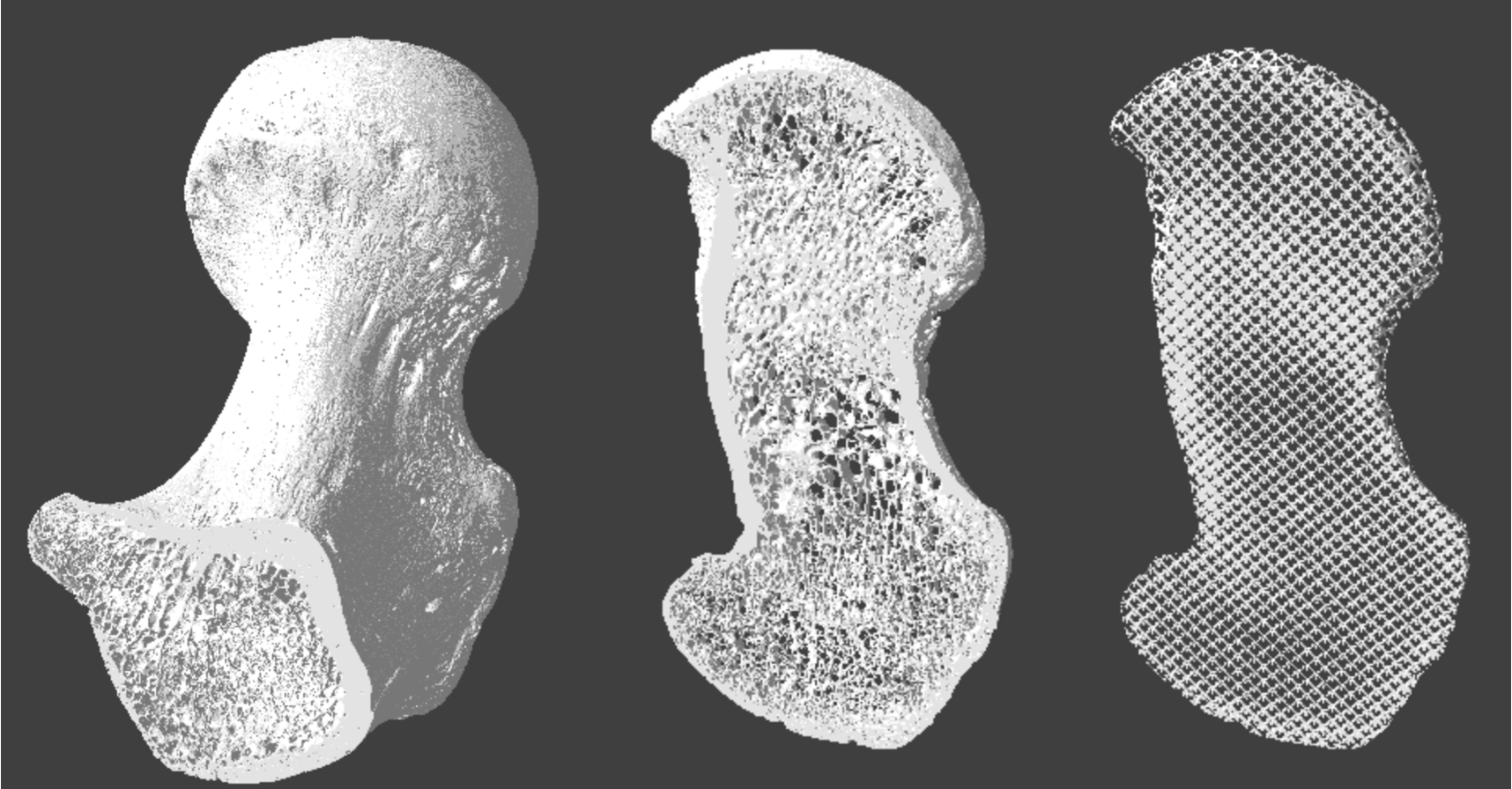}
\caption{The CT image of a human femur and its cross-section as visualized by the multiscale shape-material modeling API. Stratified Random Sampling can be used to compare microstructural statistics, such as correlation functions for bone scaffold modeling.}
\label{fig:bone}
\end{figure}

\paragraph{Stratified Random Sampling}
While truly random sampling is difficult to do on neighborhood-by-neighborhood basis, stratified random sampling lends itself well to this approach because the set of points inside a neighborhood is not dependent on the results of other neighborhoods. Like with regular grids, we simply generate each set of points as we step through the neighborhoods, randomly generating a preset number of points per neighborhood cell. The stratified random sampling is useful for downstream applications including Monte Carlo integration and computing the microstructural statistics of random heterogeneous material (Figure \ref{fig:bone}).

\subsection{Ray-Casting}
\label{sec:raycasting}

\subsubsection{Ray-Casting for Multiscale Objects}
Ray-casting is a technique for generating a 2D render of the multiscale model by shooting rays into the model and testing for intersections. For our query-based approach, this is generally handled by stepping through the multiscale model and testing for point membership at each step. Instead of redoing the process of testing membership at each scale at each step, we choose to ray-cast at each scale individually to obtain a ray entry and exit point for that scale, before attempting to ray-cast at the next-finer scale; this lets us use specialized ray-intersection queries where applicable.

Within each scale, step length is determined both by the scale's size and by taking the distance query in addition to the point-membership query; the distance can be used to increase step length while guaranteeing that we do not miss a surface. Furthermore, should a ray intersect the surface, the ray-casting is terminated for that ray. In general, each ray is handled independently in parallel with no special per-cell computation.

\subsubsection{Set Queries for Ray-Casting}

While the requirement of many point queries for each ray suggests that our set-based approach will be immediately useful to ray-casting, the variable step length and early ray termination acceleration strategies complicate application of our approach. It becomes impossible for us to pre-generate a structured point set, as not all query points will need to be tested, and testing extra points to generate cancels the performance improvement we would get from the set-based query.

We implement ray-casting by assigning each ray to a neighborhood as in the arbitrary point sampling algorithm and recomputing this at every step so long as any ray is still alive. Rays that have reached termination are tracked and are not grouped. In order to avoid constantly re-sorting rays that have not left their current neighborhood cell, we implement the ray marching algorithm into the per-neighborhood query and continue to advance each ray until it leaves the cell or intersects a surface.

While implementing the ray marching within the per-neighborhood function greatly reduces the number of times we need to do precomputation for each cell, the sequential nature of ray-casting still leads to some cells being considered more than once, such as if a ray enters a cell other rays had left in the previous step.

We alleviate this issue by exploring the use of a cache to store the per-cell precomputation, both in the context of temporarily storing this information for rays that intersect cells that had already been considered and in the context of re-using the information between render calls. This can accelerate both each individual frame, as well as consecutive frames with a moving camera. We use a simple first-in-first-out cache with limited size as a proof-of-concept.

\subsection{Slicing}

Slicing is the process of generating a stack of axis-aligned 2D images of the multiscale structure, implemented by the repeat evaluation of a point-membership query at each location for each slice. This is important for further applications such as manufacturing the multiscale structure, as each slice can then be transformed into a set of instructions for 3D printing ~\cite{liu2021memory}.

Slicing generates a regular grid of sample points which is ideal for our set-based query. For lower resolutions where the entire stack of images can fit in memory, we can directly apply our structured set-query on the input. For higher resolutions, we break the input apart into a smaller set of stacks along the height axis, outputting each after the previous output has been used, and defaulting to a stack size that aligns with the height of one neighborhood cell; this keeps us from needing to recompute the per-cell computation more than once per cell. We could also support breaking these images apart along other axes.

%
\section{Results}
\label{sec:results}

\subsection{Testing Methodology}
\label{sec:testmethod}



We run our tests on a consumer PC running an Intel i7 10700k CPU with 16 threads at 3.8 GHz and 64 GB of memory. While our approach may be extended to the GPU, for simplicity we choose to test it only with a CPU implementation. All timings use wall time; wall times for detailed tests are computed per-thread, and averaged across each thread.

We use several datasets for the performance testing of our program. We primarily use a 2-scale model with a coarse-scale polygon model and one finer scale. For the simple case, we use a simple cube for the coarse scale; different coarse scales increase the computation done at the coarse scale but reduce the amount of computation for each fine scale. For the fine-scale, we use lattices, the gyroid function, or Voronoi foams with a regular-grid neighborhood of $32^3$, $128^3$, or $256^3$ cells. We test the Voronoi foam with pre-generated edges both according to the original algorithm (Method 1) and according to our full precomputed approach (Method 2). See Section~\ref{sec:setQVF} for more details. 

The gryoid function for functional representation is produced using the libfive solid modeling library, using the ``gryoid'' function call with a period of $2\pi^2$.




\begin{figure}[htb]
\centering
\includegraphics[width=0.45\linewidth]{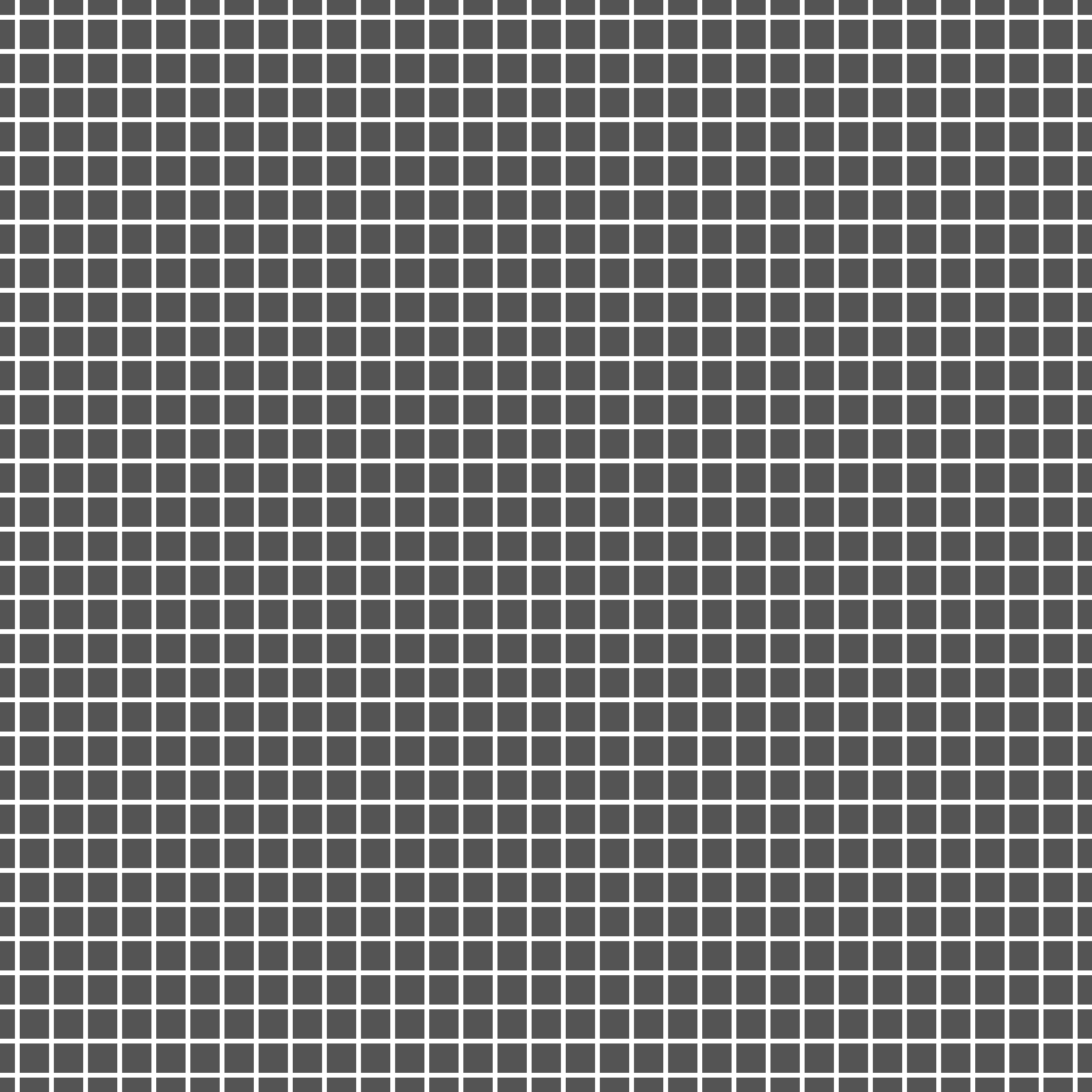}
\includegraphics[width=0.45\linewidth]{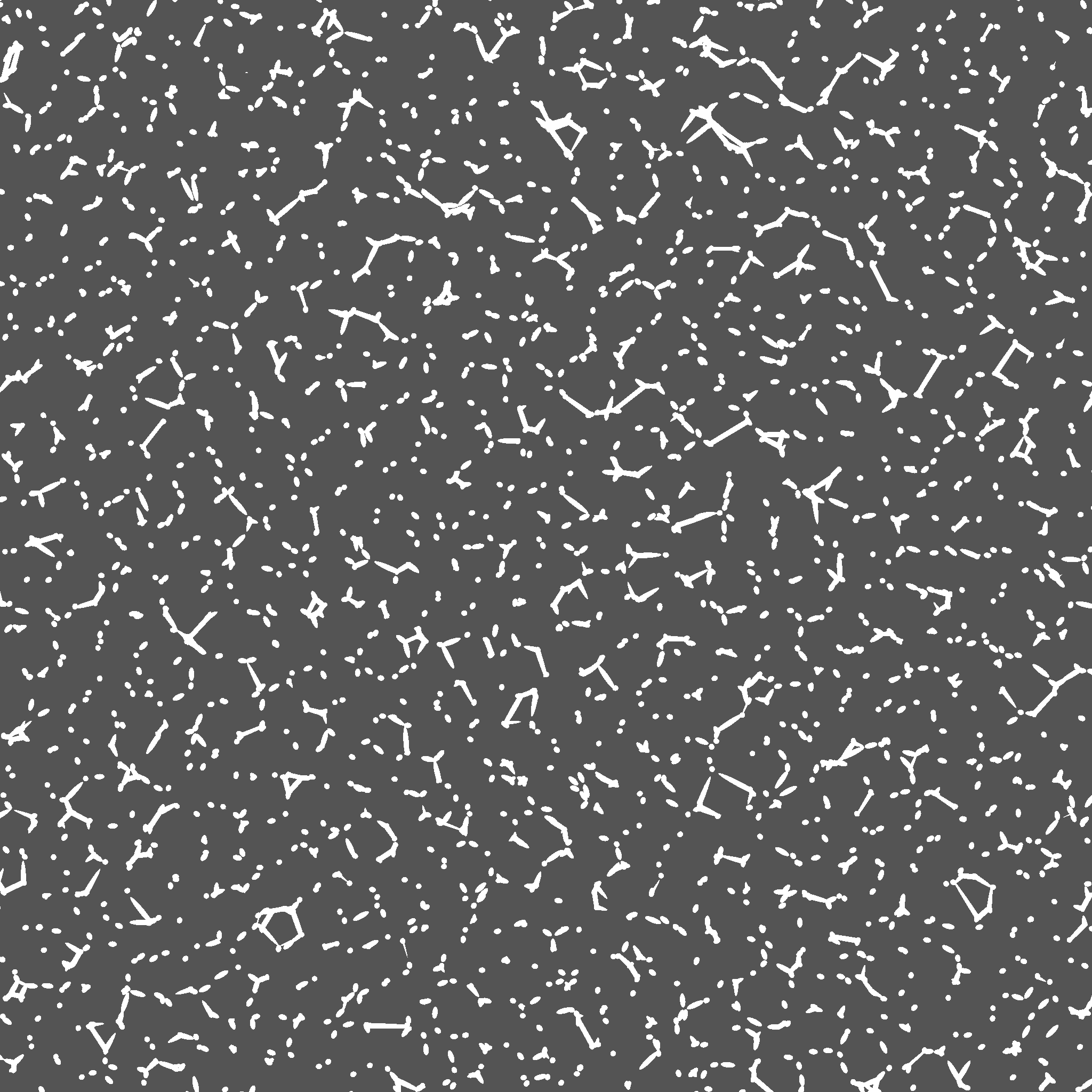}
\caption{Example slices from the parametric lattice and Voronoi foam datasets, respectively.}
 \label{fig:voronoiFoamEx}
\end{figure}

We test the slicing application in parallel, running the membership query on a $512^3$, $1024^3$, and $2048^3$ grid of sample points. For the $1024^3$ and $2048^3$ grids, we also break the slicing apart such that only one Z-layer of the neighborhood grid is computed at once, simulating the case where the entire sample grid cannot fit in memory.

We only test the ray-casting application on the Voronoi Foam microscale model with a regular-grid neighborhood of $32^3$, $128^3$, and $256^3$ cells. We sample a $256^2$, $1024^2$, and $2048^2$ image from two different angles, one in front of the model, and one from the corner with both a regular ray-casting algorithm and using our set-based query. For ray-casting, we use up to 3 steps per neighborhood grid cell, with the step length set to the distance query results according to sphere-tracing. The two methods differ only in how they sample the fine scale; sampling of the coarse scale is identical in both cases.

We also evaluate the performance of caching cells in the ray-tracing as explained in Section~\ref{sec:raycasting}. We use a simple first-in-first-out cache that stores the precomputed Voronoi edges with a maximum size of 1 GB. We provide timing results both for the situation where the image is taken with a cache and for when the frame has been cached, approximating the rendering time when the camera location is only changed by a small amount.

\subsection{Slicing Results}
\label{sec:sliceres}

\begin{table}[!t]
\caption{Slicing results on a variety of datasets and grid sizes. Timing is provided for the fine scale only, except in rows where there is no fine scale, in which we report the coarse-scale runtime. For Voronoi Foams, Set Query 1 refers to the algorithm where we only pre-compute edge position and direction, while in 2 we also pre-compute rejection of non-Voronoi edges and edge length.} 
\label{tab:sliceRes}
\centering
\begin{tabular}{|p{0.11\linewidth}| p{0.08\linewidth}| p{0.07\linewidth}| p{0.115\linewidth}|
p{0.085\linewidth}|  
p{0.085\linewidth}|  
p{0.100\linewidth}|  
}
\hline
  Coarse Model & Fine Scale & Grid Size & Method & $512^3$ & $1024^3$ & $2048^3$ \\ \hline
\hline 
  Cube    & None &$N/A$   & \naive  &$55.50$s &$453.1$s &$3399.$s \\ \hline
          &      &        & Set Q.  &$23.86$s &$200.7$s &$1578.$s \\ \hline
  Cube    & UC   &$32^3$  & \naive &$3.014$s &$5.058$s &$231.1$s \\ \hline
          &      &        & Set Q  &$25.26$s &$193.5$s &$1557.$s \\ \hline
  Cube    & UC   &$128^3$ & \naive &$1.567$s &$45.28$s &$383.3$s \\ \hline
          &      &        & Set Q  &$25.14$s &$280.2$s &$1819.$s \\ \hline
  Cube    & UC   &$256^3$ & \naive &$2.311$s &$99.22$s &$585.4$s \\ \hline
          &      &        & Set Q  &$26.54$s &$581.2$s &$2690.$s \\ \hline
  Cube    & Func &$32^3$  & \naive &$439.2$s &$3230.$s &$27208$s \\ \hline
          &      &        & Set Q  &$167.4$s &$1392.$s &$10894$s \\ \hline
  Cube    & Func &$128^3$ & \naive &$438.9$s &$3518.$s &$28240$s \\ \hline
          &      &        & Set Q  &$126.0$s &$1605.$s &$11695$s \\ \hline
  Cube    & Func &$256^3$ & \naive &$440.5$s &$3950.$s &$30038$s \\ \hline
          &      &        & Set Q  &$429.4$s &$1726.$s &$13281$s \\ \hline
  Cube    & VF   &$32^3$  & \naive  &$212.5$s &$1739.$s &$14015$s \\ \hline
          &      &        & Set Q.1 &$17.47$s &$147.7$s &$1160.$s \\ \hline
          &      &        & Set Q.2 &$21.62$s &$206.4$s &$1601.$s \\ \hline
  Cube    & VF   &$128^3$ & \naive  &$205.4$s &$1848.$s &$14327$s \\ \hline
          &      &        & Set Q.1 &$169.4$s &$542.3$s &$1984.$s \\ \hline
          &      &        & Set Q.2 &$163.2$s &$479.9$s &$2330.$s \\ \hline
  Cube    & VF   &$256^3$ & \naive  &$213.5$s &$2186.$s &$16382$s \\ \hline
          &      &        & Set Q.1 &$1117.$s &$2723.$s &$4375.$s \\ \hline
          &      &        & Set Q.2 &$1155.$s &$2756.$s &$3973.$s \\ \hline
\end{tabular} 
\end{table}

We present our overall timing results for the slicing test case in Table~\ref{tab:sliceRes}, which separately presents the times for each type of coarse and fine scale; coarse-scale times do not depend on the fine-scale type. The \naive approach refers to testing with individual point queries and set Q refers to the set query. For Voronoi Foams, set Q.1 refers to the set query with precomputed edges according to the original algorithm, and set Q.2 refers to those with fully precomputed edges.

We expect that our set-query based methods will function better in those cases where there are many sample points per cell, and will function very poorly where there are only a few sample points per cell, such as when the neighborhood and sample grid sizes are both $512^3$. We present the raw times using our algorithm even in these cases; in practice, we would switch to the \naive point-by-point approach when the number of points is too low.

We find that the timing results when testing only a single Z-layer of the layer-by-layer slicing tests are consistent among all layers tested, even for the non-repeating Voronoi Foam structure. We therefore obtain the results for these cases by averaging the times for the first three layers, and multiplying by the total number of layers.

\begin{table}[!t]
\caption{Detailed timing results. We display detailed per-cell and per-point time costs for each of the investigated approaches. There is no per-cell computation for the \naive approach. Overhead displays the amount of extra overhead per run, including costs to generate points. Points-per-cell shows the amount of points per cell required for the Set-based query to offer better performance than the \naive approach.} 
\label{tab:detailRes}
\centering
\begin{tabular}{|p{0.12\linewidth}| l|
p{0.07\linewidth}|  
p{0.13\linewidth}| 
p{0.12\linewidth}|  
p{0.10\linewidth}|  
}
\hline
  Fine Model & Method & Over-head & Per-Cell & Per-Point & Points per Cell \\ \hline
\hline 
  Voronoi Foam & \naive  & $N/A$          & $N/A$ & $3.11e^{-5}$s & $N/A$ \\ \hline
               & Set Q.1 & $4.7$s & $1.129e^{-3}$s & $1.54e^{-6}$s & $38$ \\ \hline
               & Set Q.2 & $1.9$s & $1.183e^{-3}$s & $5.72e^{-7}$s & $39$ \\ \hline
\end{tabular} 
\end{table}

We also show detailed timing results for the Voronoi Foam in Table~\ref{tab:detailRes}, which shows our computed per-cell and per-point costs for the Voronoi foam approach, as well as the critical points at which the set-based query is more efficient. The timing results are generally consistent with our expectations, with the set-based query having better relative performance where there are more sample points per grid cell. We find that the layer-by-layer slicing introduces some additional overhead costs which can be significant, especially for the set-based query. 

We find that our grouping improves the performance for the coarse scale, for which we did not apply any other optimizations and which we still test point-by-point within each group. We believe that this result stems from the improved data locality of sample points which are organized within groups. 

We find that our set query with specialized precomputation for lattice structures does not improve the timing results, likely due to the extreme simplicity of the parametric lattices which we tested. A more complex repetitive structure, the gyroid function, benefits much more significantly from the same approach.

We find that both our approaches to the Voronoi Foam have similar costs on a per-cell basis, with Method 1 being slightly faster per-cell, while Method 2 is much faster per-sample point (See Section~\ref{sec:testmethod} for details). 
Both approaches are more efficient than the \naive method even with much fewer points-per-cell than with Unit Cells; however, where there are very small amounts of points per cell, the performance is significantly worse with the set-based query. As mentioned previously, in practice we can switch to the individual point query where the amount of points per cell is below the inflection point, preventing this from being a problem.

From the detailed timing results for Voronoi foams, we expect that approximately 40 points per cell are necessary for the set-based query to be more efficient, or a sample grid at least $3.4$x larger per side than the neighborhood grid. Our slicing test results generally support this except for one outlier due to the extra overhead from layer-by-layer slicing.



\subsection{Rendering Results}

\begin{table*}[!t]
\caption{Rendering results on a variety of datasets and grid sizes. Timing is provided for the fine scale only, except in rows where there is no fine scale, in which we report the coarse-scale runtime. Set Query 1 refers to the algorithm where we only pre-compute edge position and direction, while in 2 we also pre-compute rejection of non-Voronoi edges and edge length.} 
\label{tab:renderRes}
\centering
\begin{tabular}{|p{0.06\linewidth}| p{0.06\linewidth}| p{0.04\linewidth}| p{0.06\linewidth}| p{0.06\linewidth}| p{0.06\linewidth}|
p{0.085\linewidth}|  
p{0.085\linewidth}|  
p{0.09\linewidth}|  
}
\hline
  Coarse Model & Fine Scale & Grid Size & Method & View & Cached? & $512^3$ & $1024^3$ & $2048^3$ \\ \hline
\hline 
  Cube    & None &$N/A$   & \naive  & Front  & N &$0.153$s &$0.677$s &$2.739$s \\ \hline
          &      &        &         & Corner & N &$0.086$s &$0.764$s &$2.785$s \\ \hline
          &      &        & Set Q.  & Front  & N &$0.050$s &$0.710$s &$2.621$s \\ \hline
          &      &        &         & Corner & N &$0.048$s &$0.730$s &$2.796$s \\ \hline
  Cube    & VF   &$32^3$  & \naive  & Front  & N &$2.960$s &$52.35$s &$206.5$s \\ \hline
          &      &         &         & Corner & N &$3.213$s &$52.52$s &$208.6$s \\ \hline
          &      &        & Set Q.1 & Front  & N &$1.454$s &$4.172$s &$11.21$s \\ \hline
          &      &        &         & Front  & Y &$0.251$s &$2.252$s &$9.024$s \\ \hline
          &      &        &         & Corner & N &$1.594$s &$4.484$s &$12.29$s \\ \hline
          &      &        &         & Corner & Y &$0.220$s &$2.585$s &$10.31$s \\ \hline
          &      &        & Set Q.2 & Front  & N &$1.343$s &$3.083$s &$6.699$s \\ \hline
          &      &        &         & Front  & Y &$0.138$s &$1.168$s &$4.591$s \\ \hline
          &      &        &         & Corner & N &$1.530$s &$3.279$s &$7.650$s \\ \hline
          &      &        &         & Corner & Y &$0.139$s &$1.410$s &$5.711$s \\ \hline
  Cube    & VF   &$128^3$ & \naive  & Front  & N &$3.467$s &$59.29$s &$240.4$s \\ \hline
          &      &        &         & Corner & N &$3.639$s &$59.55$s &$241.5$s \\ \hline
          &      &        & Set Q.1 & Front  & N &$10.52$s &$37.31$s &$61.35$s \\ \hline
          &      &        &         & Front  & Y &$9.523$s &$40.05$s &$61.49$s \\ \hline
          &      &        &         & Corner & N &$13.22$s &$59.79$s &$127.6$s \\ \hline
          &      &        &         & Corner & Y &$12.56$s &$61.52$s &$127.5$s \\ \hline
          &      &        & Set Q.2 & Front  & N &$9.212$s &$25.35$s &$49.57$s \\ \hline
          &      &        &         & Front  & Y &$1.912$s &$25.09$s &$38.42$s \\ \hline
          &      &        &         & Corner & N &$9.214$s &$36.38$s &$61.17$s \\ \hline
          &      &        &         & Corner & Y &$1.914$s &$29.19$s &$44.56$s \\ \hline
  Cube    & VF   &$256^3$ & \naive  & Front  & N &$3.601$s &$61.13$s &$247.0$s \\ \hline
          &      &        &         & Corner & N &$3.796$s &$61.23$s &$245.8$s \\ \hline
          &      &        & Set Q.1 & Front  & N &$26.48$s &$133.6$s &$619.8$s \\ \hline
          &      &        &         & Front  & Y &$26.01$s &$133.4$s &$619.2$s \\ \hline
          &      &        &         & Corner & N &$31.56$s &$218.1$s &$420.0$s \\ \hline
          &      &        &         & Corner & Y &$31.45$s &$218.1$s &$421.3$s \\ \hline
          &      &        & Set Q.2 & Front  & N &$24.60$s &$96.92$s &$531.6$s \\ \hline
          &      &        &         & Front  & Y &$14.94$s &$91.32$s &$531.5$s \\ \hline
          &      &        &         & Corner & N &$29.53$s &$118.1$s &$197.8$s \\ \hline
          &      &        &         & Corner & Y &$17.29$s &$117.0$s &$209.2$s \\ \hline
\end{tabular} 
\end{table*}

We present our overall timing results for the rendering test cases in Table~\ref{tab:renderRes}, which separately presents the times for each type of coarse and fine scale; coarse-scale times do not depend on the fine-scale type. Again, the \naive approach refers to testing with individual point queries, set Q.1 refers to the Voronoi Foam set query with precomputed edges according to the original algorithm, and set Q.2 refers to those with fully precomputed edges.

We find that where there is a sufficient amount of sample points per-cell, our approach is significantly faster compared to the \naive method; however, while the \naive approach barely increases in cost with an increase in the number of neighborhood cells, our approaches are significantly affected and eventually become slower.

Likely due to the larger amount of samples here, we find that Method 2 is significantly faster for the rendering scenario; furthermore, it benefits significantly from caching, since much of the per-sample point work is pushed into per-cell precomputation. 

Our cache will eventually become saturated with a larger neighborhood grid size but generally isn't significantly slower even when full. We used an arbitrary 1 GB cache for our test cases to provide results where this occurs; in practice, a larger cache can easily be used for CPU rendering.

\section{Conclusion}

We have proposed a set-based approach to the query-based methods used for multi-scale shape-material modeling. It retains the compatibility and usability of the point-based queries with the performance that can be achieved through precomputation, while avoiding precomputation over the whole domain, which would be intractable for very fine structures. 

Our approach does have some limitations; one of the largest is that the points that we are sampling must be known ahead of time--that is, the sample location cannot depend on the results of existing points and be tested simultaneously. 

We also rely on the exploitation of some form of repeated or common information for each type of microscale function for our maximal benefits. In general, how this is done depends directly on the microscale function, and therefore requires anyone adding a new representation to write their set-based query methods to gain the performance benefits--however, by defaulting to a \naive{} point-based, we are still able to use arbitrary representations without any additional work by the authors, albeit with less performance than for customized implementations.

We have also only tested our approach with local fine-scale structures; parametric lattices only require information from the current neighbourhood cell, and Voronoi foams require information from the current cell, it's neighbours, and the neighbours of those cells. Our approach has not been tested with fully global fine-scale approaches.

Our performance benefits assume that we have more sample points and more neighborhood cells in the microscale than we have simultaneous threads to work with; if these conditions are not true, such as on a large-scale cluster machine, it could be more efficient to sample each point simultaneously without precomputation. However, if we consider a machine with more CPU cores than neighborhood cells, it implies that we likely have enough memory to store the precomputed information on a per-cell basis across multiple queries as well.

The simplest direction for future work is to come up with set-based query optimizations for other types of microscale representations, such as those based on procedural textures. We could also seek to implement these methods on the GPU to consider a situation with significantly more parallelism. Other avenues include developing algorithms which group points more efficiently; finding specialized algorithms for specific, common set queries; or testing whether the consistent grouping of sample points is more efficient when many different scales are considered at once.

\section*{Acknowledgements}
This research was supported by the Defense Advanced Research Projects Agency. The responsibility for errors and omissions lies solely with the author. 

\section*{References}

\bibliography{mybibfile}

\end{document}